\documentclass[10pt,twocolumn,letterpaper]{article}

\usepackage[pagenumbers]{cvpr}
\usepackage{amsmath,amssymb}
\usepackage{booktabs}
\usepackage{array}
\usepackage{multirow}
\usepackage{tabularx}
\usepackage{makecell}
\usepackage{xspace}
\usepackage{microtype}
\usepackage{tikz}
\usetikzlibrary{arrows.meta,calc,fit,positioning,shapes.geometric}

\newcommand{\system}{ChemXRG\xspace}

\newcommand{\code}[1]{\texttt{#1}}

\definecolor{chemblue}{HTML}{2F6FDB}
\definecolor{chemgold}{HTML}{D99A16}
\definecolor{chemgreen}{HTML}{168B64}
\definecolor{chemred}{HTML}{C94848}
\definecolor{chempurple}{HTML}{7656A5}
\definecolor{chemlight}{HTML}{F2F5F8}

\definecolor{cvprblue}{rgb}{0.21,0.49,0.74}
\usepackage[breaklinks,colorlinks,allcolors=cvprblue]{hyperref}
\hypersetup{pdfauthor={Hongyuan Wang, Daming Luo, Nico Pietroni, Christy Liang, Andrew McDonagh},
pdftitle={Constrained Program Generation for 3D Reaction Animation with a 0.8B Model}}

\title{Constrained Program Generation for 3D Reaction Animation with a 0.8B Model}

\author{Hongyuan Wang\thanks{Equal contribution. Hongyuan Wang and Daming Luo are co-first authors.}\quad
Daming Luo\footnotemark[1]\\[3pt]
Nico Pietroni\quad Christy Liang\quad Andrew McDonagh\\[6pt]
University of Technology Sydney, Australia}

\begin{document}
\maketitle
\begin{figure*}[t]
\centering
\includegraphics[width=.98\textwidth,height=.29\textheight,keepaspectratio]{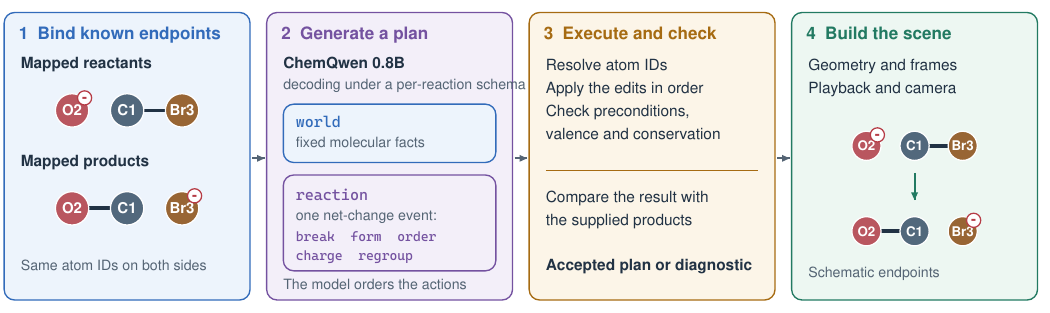}
\caption{\textbf{The ChemXRG framework.}
A mapped reaction passes through input binding (blue), DSL generation (purple), execution checks (ochre) and visualization (green).
The same atom identifiers connect the supplied endpoints to the plan and its displayed change; input-derived facts constrain generation, while the model selects the event's actions.
Molecular drawings are schematic, with hydrogens omitted; they illustrate the representation rather than a sampled trajectory.
\Cref{fig:program} expands the SN2 program.}
\label{fig:teaser}
\end{figure*}

\begin{abstract}
Visualizing a chemical reaction requires making its molecular changes visible while keeping the animation faithful to the stated chemistry.
Equations, structural diagrams and molecular viewers provide complementary descriptions, but assembling an interactive three-dimensional explanation still requires specifying the changes and checking their consistency.
We present \system, a domain-specific language (DSL) framework that addresses this gap by representing a reaction animation as an executable program.
Persistent atom identifiers and explicit bond, charge and grouping operations connect the symbolic reaction to the displayed transformation.
This shared representation lets generation, verification and rendering operate on the same account of what changes.

Given known, atom-mapped reactant and product structures, reaction-grounded constraints fix input-determined facts and restrict action choices; execution checks validate the resulting transformation before geometry and frames are constructed.
We implement this paradigm with a reaction-program corpus and ChemQwen, a trained 0.8B DSL generator.
Paired and component evaluations show improved compiler acceptance and normalized full-program agreement under input-conditioned constraints, while identifying remaining failures that require execution checks.
A public browser application demonstrates the connection from symbolic reaction descriptions to inspectable programs and interactive 3D animations.
\end{abstract}

\section{Introduction}
\label{sec:intro}

For people to understand a chemical reaction, they must connect a symbolic equation to changes in molecular structure.
On the page, reactants and products can be drawn side by side, but the reader must still connect atom identities, bond changes and their spatial presentation.
In an SN2 example, this means following the same carbon as a leaving group departs and a nucleophile becomes attached.
Interactive animation can expose these relationships through playback and spatial inspection, motivating its use in chemical communication and teaching~\cite{Tasker2006Visualisation}.
The challenge is to construct such a display systematically while ensuring that the molecules and transformations shown agree with the supplied reaction.

Existing representations solve different parts of this problem.
Equations communicate composition and stoichiometry, while two-dimensional structural diagrams and reaction schemes depict connectivity and selected transformations.
Machine-readable encodings such as SMILES and atom mapping make molecular graphs and endpoint correspondences available to computation~\cite{Weininger1988SMILES,Schwaller2021RXNMapper}; viewers such as 3Dmol.js and Mol* render supplied structures and trajectories~\cite{Rego2015ThreeDmol,Sehnal2021Molstar}.
Even with both endpoints available, an animation needs an explicit account of which atoms persist, which operations occur and how they are presented.
Producing this event description and verifying its correspondence to the reaction remains a separate authoring task.

Symbolic generation suggests how to bridge this gap.
LaMoGen~\cite{jiangLaMoGenLanguageMotion2026}, for example, lets an LLM compose LabanLite motion symbols and uses a learned decoder to turn them into human motion.
An explicit intermediate language connects semantic intent to visual realization, but reaction animation also requires that the resulting program remain grounded in the supplied molecules.
A well-formed description can still insert an extra carbon or assign the wrong charge, changing the substance being depicted.
The central requirement is therefore a representation that makes chemical changes executable and checkable while preserving the identities established by the input.

We propose \system, a DSL-based framework that connects reaction notation, executable transformations and interactive 3D animation (\Cref{fig:teaser}).
The DSL gives each atom a persistent identifier and represents the displayed change as explicit bond, charge and grouping operations.
These operations provide a common description for the generator, execution verifier and renderer, so the animation can be inspected together with the program that specifies it.
Reaction-grounded constraint compilation fixes known input facts and restricts the actions available during generation.
The verifier executes the plan and checks its agreement with the supplied endpoints; the materializer then constructs geometry and animation frames.

To evaluate this paradigm, we construct a reaction-program corpus and train ChemQwen, a 0.8B model that generates DSL plans from mapped endpoints.
Paired and component studies show how input-conditioned constraints improve program acceptance and consistency, and where execution checks remain necessary; \Cref{sec:experiments} reports the quantitative results and failure analysis.
Our contributions are:
\begin{itemize}[leftmargin=*,nosep]
\item An executable reaction-animation language that connects static molecular descriptions to explicit changes, persistent atom identities and interactive 3D rendering (\Cref{sec:representation}).
\item Input-conditioned generation and execution checks that bind known molecular facts and validate the generated transformation against the supplied reaction (\Cref{sec:method}).
\item A reaction-program corpus, a trained compact generator and a public browser application that demonstrate this language as an interface between chemical notation and animation (\Cref{sec:data,sec:experiments,sec:runtime}).
\end{itemize}
We study animation of a single net structural change between known, mapped endpoints; the generated plan specifies the change to display, rather than predicting an unknown product or a physical reaction trajectory.

\section{Related Work}
\label{sec:related}

Efforts to make chemical reactions understandable have developed along complementary lines: notation for communicating chemical change, encodings for computation, and visual tools for inspecting molecular structure and motion.
Handwritten formulas and two-dimensional reaction schemes provide a human-readable language; line notations such as SMILES later made molecular structures convenient for machine processing~\cite{Weininger1988SMILES}.
Interactive graphics and molecular animations added spatial inspection and playback~\cite{Humphrey1996VMD,Tasker2006Visualisation}.
These approaches coexist, and \system connects their roles through a DSL that makes a particular reaction's displayed transformation executable and checkable, with a learned generator supporting program construction.

\paragraph{Chemical notation and reaction data.}
Written formulas and reaction equations compactly record composition and stoichiometry, while two-dimensional structural drawings and reaction schemes make connectivity and selected changes visible.
Their interpretation still requires connecting symbols to molecular structures and processes~\cite{Kozma1997Multimedia}.
For computation, Weininger introduced SMILES in 1988~\cite{Weininger1988SMILES}, encoding molecular graphs as strings of atoms, bonds, branches and ring closures.
Reaction SMILES combines endpoint structures, and atom mapping adds correspondences across them.
Krenn et al.'s SELFIES (2020)~\cite{Krenn2020SELFIES} uses semantic derivation rules to enforce molecular validity, while Schwaller et al.'s RXNMapper (2021)~\cite{Schwaller2021RXNMapper} extracts atom correspondences from attention in a model trained on reaction strings.
These advances improve machine readability, validity and correspondence; an animation additionally needs executable changes and a playback description.

\paragraph{Learning from reactions.}
Reaction prediction extends structural representation toward inferring chemical outcomes from data.
Jin et al. (2017)~\cite{Jin2017WLDN} identify a reaction center with a Weisfeiler--Lehman network, enumerate candidate products through graph edits and rank them with a difference network.
Schwaller et al.'s Molecular Transformer (2019)~\cite{Schwaller2019MolecularTransformer} instead uses attention to translate reactant and reagent SMILES into product SMILES.
Graph edits explicitly describe structural change, whereas sequence translation learns the mapping between endpoint strings.
Both address product prediction; an inferred product alone does not specify an animated scene, persistent visual entities or playback.
Our setting supplies both endpoints and uses the DSL to express, execute and check the transformation to be displayed.

\paragraph{Molecular visualization.}
Molecular visualization has long supported interactive inspection of spatial structures and trajectories.
Humphrey, Dalke and Schulten presented VMD in 1996~\cite{Humphrey1996VMD} for molecular graphics and trajectory analysis; Hanson described Jmol's interactive crystallographic visualization in 2010~\cite{Hanson2010Jmol}.
Rego and Koes introduced 3Dmol.js in 2015~\cite{Rego2015ThreeDmol}, exposing molecular rendering through browser WebGL.
More recent systems include Pettersen et al.'s ChimeraX (2021)~\cite{Pettersen2021ChimeraX} and Sehnal et al.'s Mol* (2021)~\cite{Sehnal2021Molstar}, which support rich molecular scenes, with Mol* also providing playback of supplied trajectories.
These tools make given molecular data explorable, but the changes and scene instructions for a particular reaction still have to be supplied.
Our DSL makes this event description executable, checkable and available to the visualization runtime.

\paragraph{Visual programs and structured generation.}
Structured visual generation makes the link between a requested change and its realization explicit.
Sun et al.'s LayoutVLM (2024)~\cite{sunLayoutVLMDifferentiableOptimization2024} predicts both initial object poses and spatial relations, then refines the layout through differentiable optimization.
Jiang et al.'s LaMoGen (2026)~\cite{jiangLaMoGenLanguageMotion2026} uses retrieval-augmented prompting to compose LabanLite symbols with body-part and temporal attributes; a detail augmentor enriches the plan, and a jointly trained codebook and Transformer decoder reconstruct motion.
Both separate semantic specification from low-level realization, giving the intermediate representation a central role.
For reaction animation, that representation must additionally define bond and charge operations over persistent atoms and support comparison with the given product graph.
\system instantiates this separation with an executable chemical language, input-derived constraints and graph-transition checks before materialization.

\paragraph{Constrained decoding.}
An executable representation still leaves the risk of generating invalid or inconsistent programs.
Scholak et al.'s PICARD (2021)~\cite{Scholak2021PICARD} rejects inadmissible tokens through incremental parsing, while Poesia et al.'s Synchromesh (2022)~\cite{Poesia2022Synchromesh} uses completion engines to enforce syntax, typing and contextual constraints.
Dong et al.'s XGrammar (2024)~\cite{dongXGrammarFlexibleEfficient2024} accelerates grammar-based token filtering by combining precomputed checks with runtime parsing.
Applying these mechanisms to reaction animation requires deciding which facts the input fixes and which properties depend on executing the selected actions.
A generic output schema alone does not bind a program to a particular molecule, and static restrictions on individual actions do not necessarily prevent conflicts across actions.
We derive the schema from each mapped reaction and complement decoding with execution checks, separating input-preserving construction from validation of the generated transformation.

\paragraph{Visual explanations in chemistry.}
Animation is useful only if viewers can connect what moves to what the chemical symbols mean.
Kozma and Russell (1997)~\cite{Kozma1997Multimedia} studied expert and novice interpretations of chemical representations, and Wu, Krajcik and Soloway (2001)~\cite{Wu2001Promoting} examined students' use of a visualization tool to develop representational understanding.
Tasker and Dalton (2006)~\cite{Tasker2006Visualisation} discussed molecular animation design for learning, while Tversky et al. (2002)~\cite{Tversky2002Animation} cautioned that animation alone does not ensure better understanding.
Our contribution addresses the generation and inspection of the representation: mapped atoms retain their identifiers between input, program and scene; measuring educational benefit requires a separate user study.

\section{The ChemXRG Domain-Specific Language}
\label{sec:representation}

\begin{figure*}[t]
\centering
\includegraphics[width=.98\textwidth,height=.28\textheight,keepaspectratio]{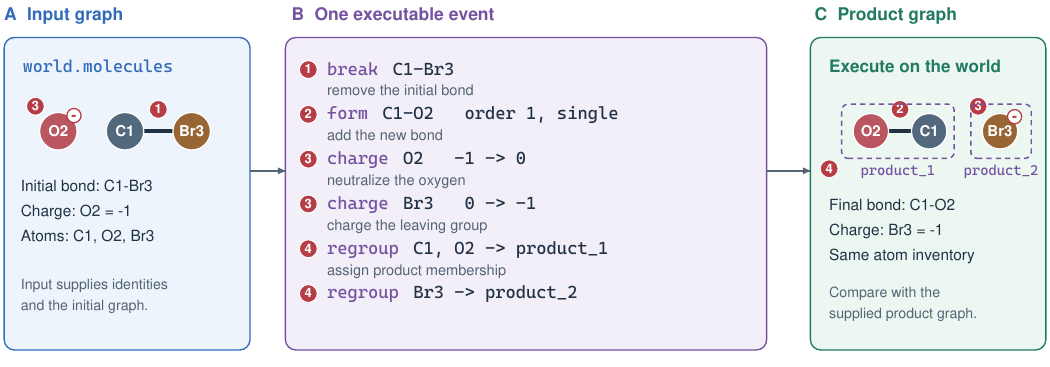}
\caption{\textbf{A mapped SN2 reaction in the ChemXRG DSL.}
The initial graph (A), abbreviated action list (B) and executed product graph (C) share persistent atom identifiers.
Bond removal, bond formation, charge updates and regrouping specify one net-change event; the executed graph is compared with the supplied product graph, and the numbered discs link each action to the bond, charge or group it changes.
The action notation is abbreviated for readability, and the drawings omit hydrogens and geometry; neither intermediate physical states nor measured trajectories are implied.}
\label{fig:program}
\end{figure*}

The representation is the central design decision of this work, because everything downstream, the corpus, the learned generator, the decoding constraints and the checks, operates on it.
We had four requirements for it.
It must preserve atom identity, so that the same atom can be followed from the reactant equation through the program to every rendered frame.
It must make change explicit and executable: a plan should list the graph edits that turn reactants into products, in a form a compiler can run, rather than describe the result or a trajectory.
It must be checkable against the input, so that a wrong program is detected by execution instead of by eye.
And it must be small enough to be a learning target for a compact model, which argues against free-form animation code and against numeric coordinate trajectories; in our early trials, directly generated coordinate programs were too unreliable to build on (\Cref{sec:data}), so the language separates symbolic generation from geometric materialization.

ChemXRG therefore sits between chemical notation and a program.
Like a reaction equation it names atoms and states what is present before and after; like a program it lists operations with defined semantics and can be executed.
Geometry is deliberately outside the language: a deterministic materializer embeds reactant coordinates, relaxes the product geometry from them and adds display hydrogens, so identity persists on screen without the model writing a single coordinate.

\subsection{Atom Identity and Plan Structure}

Every mapped atom is named by its element and map number, such as \code{C1} or \code{Br3}, so the same name denotes the same atom in the reactant graph $G_R$, in the product graph $G_P$ and in every frame.
A bond is a record between two such names with an integer order and a type, and aromatic systems are written in Kekul\'e form so that every bond has a definite order.
These names are the vocabulary shared by the input, the program and the scene, and they make a plan checkable: a plan may refer only to mapped atoms, and no mapped atom may appear or disappear.
Hydrogens that take part in the event are mapped atoms like any other; the remaining hydrogens are implicit in the program and are added by the materializer for display.

A plan is a JSON document with two parts, the scene before anything happens and what happens:
\begin{itemize}[leftmargin=*,nosep]
\item \textbf{World:} \code{world.molecules} lists the reactants with their atoms, formal charges, bonds and formulas.
It is the initial graph on which the actions execute and the set of persistent entities that the viewer draws.
\item \textbf{Reaction:} \code{reaction} holds the equation, the reactant and product terms, and a \code{timeline} with one net-change event.
The event carries a duration, an optional label and narration, focus targets in \code{scaffold}, and a list of actions of five kinds: \code{break} removes a bond, \code{form} adds one, \code{order} changes a bond order, \code{atom.set\_formal\_charge} updates a charge, and \code{regroup} assigns atoms to a product.
\end{itemize}

Five action kinds cover the changes a mapped reaction exhibits at the graph level in the evaluated profile, bond topology, bond order and formal charge, plus the regrouping of atoms into products that the display needs.
Each action is a record over atom names, so it is both human-readable and directly executable; duration, label, narration and \code{scaffold} focus control the presentation and leave the molecular graph unchanged.
The evaluated profile places all actions in a single event, the net change.

\subsection{Execution Semantics}

Executing a plan is a graph transition.
Let $G_t=(V,E_t,q_t,g_t)$ be the mapped atoms, their bonds, formal charges and product membership at an event boundary; the event's action list $A_t$ maps it to
\begin{equation}
  G_{t+1}=\mathcal{T}(G_t,A_t),
\end{equation}
where $\mathcal{T}$ applies the actions in list order and is defined only when every action refers to a mapped atom and satisfies its precondition on the current state (a bond to break must exist, a bond to form must not, an order change must change the order), no two actions write the same bond, charge or group, and the result respects the valence policy and conserves the mapped atoms.
This definition is what the compiler implements: it accepts a plan when $\mathcal{T}$ is defined on it, and a separate endpoint check compares the bonds, charges and product membership of $\mathcal{T}(G_R,A)$ with $G_P$, so acceptance and endpoint agreement are distinct verdicts, and \Cref{sec:experiments} reports them separately.
Because the semantics is defined on named atoms, execution is deterministic and cheap, and the animation interpolates the display around the discrete update.

\subsection{What the Input Determines and What the Plan Adds}

Given a mapped reaction, the endpoints determine most of a plan and, through the graph difference, the net edit set itself: the initial world is $G_R$, the formulas and reactant terms follow from it, the product terms and each atom's regroup target follow from $G_P$, the atoms whose charge changes are those whose charge differs between the two, and the bonds that must break, form or change order are the edges on which $G_R$ and $G_P$ disagree.
A deterministic projector can therefore write a valid plan from the endpoints alone, and \Cref{sec:experiments} uses one as a reference point.
What a plan adds is how that change is expressed and staged: which action encoding realizes it, in what order the actions are listed, and how the event is labeled, timed and focused.
The method uses this determinacy in two places.
The generator is trained on complete plans, so it learns to write the determined fields as well as the rest (\Cref{sec:learning}); at inference, the determined fields are compiled into decoding constraints that fix the world, the derived fields, the product membership and the charge targets, and that enumerate the admissible actions without fixing the action list (\Cref{sec:method}).
The model chooses among the admissible actions and orders them, and execution then confirms that the chosen list reaches $G_P$.

\subsection{Worked Example}

\Cref{fig:program} shows the plan for the SN2 reaction \code{[CH3:1][Br:3].[OH-:2]} $\rightarrow$ \code{[CH3:1][OH:2].[Br-:3]}.
The input fixes three mapped atoms, \code{C1}, \code{O2} and \code{Br3}, their initial molecules and their product membership; the plan's one event breaks \code{C1-Br3}, forms \code{C1-O2}, neutralizes the oxygen, charges the bromine and regroups the atoms into the two products.
Executing these edits on $G_R$ yields $G_P$, so the plan is correct.

\section{Data Construction and Model Selection}
\label{sec:data}

\subsection{From Reaction Records to Program Targets}

Our early attempt to generate programs with explicit coordinates yielded only one parseable output in a 19-case pilot with a 122B Int4 LoRA model.
This motivated a separation of symbolic supervision from geometric materialization.
We construct targets from substrate tables and bond and charge edit lists: mapped atoms define the initial world, edits define the net-change event, and product membership defines regroup actions.
The resulting 34,294 ChemXRG programs are compiler-screened and reviewed during curation, covering substitution, condensation, addition, acid--base, olefination, rearrangement, cycloaddition, oxidation--reduction, hydrolysis and hydration, and elimination.
Coordinates and display hydrogens are produced by the materializer rather than included as prediction targets.

The training record partitions this corpus into 24,004 training, 3,430 validation and 6,860 test-v1 cases; the actual examples consumed by each training stage are specified below.
The decoding studies use development-200, external validation-3,698 and test-14,800, and a later pilot-200/remaining-19,028 split.
The external validation/test pair and the later pilot/remainder pair are disjoint by source identifier, recorded group and exact input.
These are distinct evaluation sets, and historical model explorations are not treated as a single controlled ablation.

\subsection{Selecting Teacher and Student Models}

Model scale and code pretraining alone did not yield a reliable generator: an independently fine-tuned 80B code model compiled only 18 of 200 development outputs in its historical configuration.
Earlier teacher pilots also explored GPT-OSS pseudo-labels and DeepSeek prompting with retrieved examples and compiler feedback.
We retained Qwen3.8-27B~\cite{Qwen38ModelCard} as the trainable teacher and Qwen3.5-0.8B~\cite{qwenteamQwen3508BModelCard2026} as the compact student, then evaluated the learned program mapping explicitly.
These choices define the subsequent training lineage; the exploratory runs changed data, representations and training settings and therefore do not establish a general model-size ranking.

\section{Supervised Learning and RL Exploration}
\label{sec:learning}

\begin{figure*}[t]
\centering
\includegraphics[width=.98\textwidth,height=.30\textheight,keepaspectratio]{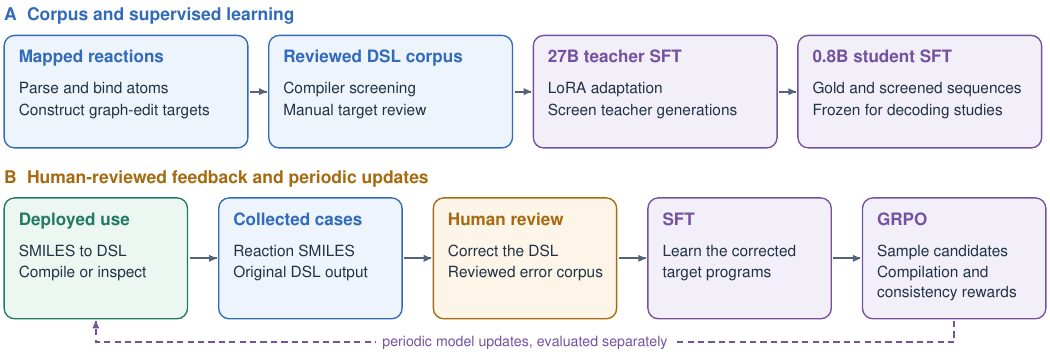}
\caption{\textbf{Learning from constructed programs and reviewed errors.}
(A) Mapped reactions become checked DSL targets; teacher SFT and screened teacher outputs support student training.
The retained student checkpoint is frozen for the decoding comparisons.
(B) Compilation failures or errors found after compilation send the input SMILES and original DSL to a cloud-hosted PostgreSQL database.
Human corrections supply targets for periodic SFT followed by GRPO; the dashed return denotes this update workflow, whose improvement is evaluated separately.}
\label{fig:learning-feedback}
\end{figure*}

\subsection{Supervised Adaptation of the Teacher}

An early 27B configuration produced 198 parseable but no compilable outputs on 200 cases, showing that learning JSON form did not suffice for the execution contract.
With the symbolic targets above, we fine-tuned the Qwen3.8-27B teacher using LoRA~\cite{Hu2022LoRA}: formal SFT consumed 10,616 examples over 2,654 steps, followed by a revised continuation on 17,852 rows over 4,463 steps.
The continuation started from merged SFT weights with fresh LoRA and optimizer state, without inheriting a failed RL candidate.
This SFT-v3.1 teacher supplied the student supervision.

\subsection{Student Fine-Tuning and the Label Control}

The teacher still produced rejected programs, so its outputs were screened before becoming student targets.
Of 4,096 teacher generations on training inputs, 3,904 survived compiler screening.
We fine-tuned all student parameters in BF16 on 17,852 gold programs plus these 3,904 teacher sequences, giving 21,756 training positions over 5,439 steps.
A matched \code{gold-control} used gold targets for the same additional inputs, preserving the base model, input order and training budget; the untrained base provided a third control.
Sequence distillation here transfers supervision, not the teacher's weights.

\begin{table}[t]
\centering
\caption{Learning configurations under unconstrained greedy decoding.
$N$ identifies the case set: validation-3,430 or test-v1-6,860 in the upper rows, development-200 in the lower.
Compiler verdicts are archived; full and reaction matches retain array order.
The lower panel compares the two matched student targets with the untrained base.}
\label{tab:learning}
\small
\begin{tabular}{lrrrrr}
\toprule
Model & $N$ & Parse & Comp. & Full & React. \\
\midrule
27B teacher & 3,430 & 3,430 & 3,165 & 3,034 & 3,046 \\
27B teacher & 6,860 & 6,860 & 6,412 & 6,204 & 6,239 \\
0.8B distilled & 6,860 & 6,858 & 6,639 & 6,475 & 6,496 \\
\midrule
0.8B gold-control & 200 & 200 & 173 & 159 & 161 \\
0.8B distilled & 200 & 200 & 173 & 152 & 154 \\
0.8B base & 200 & 14 & 0 & 0 & 0 \\
\bottomrule
\end{tabular}
\end{table}

Teacher labels did not outperform matched gold labels on development-200: both students compiled 173 programs, and gold labels gave more reference matches (\Cref{tab:learning}).
On the shared 6,860 test inputs, however, the distilled student compiled 6,639 programs against the teacher's 6,412, supporting the feasibility of the learned mapping at 0.8B scale without isolating a causal benefit from distillation.
This \code{distilled} checkpoint was frozen for the subsequent decoding studies.

\subsection{GRPO and Continual Refinement}

Execution feedback provides a way to refine DSL generation beyond fitting supervised targets.
We explored GRPO~\cite{Shao2024DeepSeekMath}, which samples multiple programs for the same reaction, compares their rewards within the group, and updates the policy with a clipped objective and reference-policy regularization.
Our reward design gives priority to compiler acceptance and incorporates input preservation, conservation and agreement with the reference program.
This connects the learning objective to properties that the generated program must satisfy when executed.

Early trials highlighted the importance of informative reward differences when several candidates fail compilation, motivating richer consistency signals and candidate resampling.
An invariant-GRPO pilot compared on-policy resampling with an off-policy expert-injection variant on the same 32 training prompts, retaining four candidates per prompt.
The latter introduced a frozen gold completion when all sampled candidates failed, providing a reference candidate within the group.
We also explored repair supervision followed by GRPO to combine explicit corrections with execution feedback.
These pilots did not establish a consistent gain over the retained SFT model, so the subsequent decoding comparisons use the frozen student checkpoint.

Beyond these pilots, we have established a human-reviewed feedback loop that connects deployed use to subsequent model updates (\Cref{fig:learning-feedback}).
When a generated DSL program fails compilation, or an error is identified after compilation, the system sends the user's reaction SMILES and the corresponding original DSL output to a PostgreSQL database hosted on a cloud server.
Periodic human review examines the recorded cases and produces corrected outputs in the standard DSL format, building an error-focused corpus of reaction inputs and reviewed target programs.

The update workflow uses these reviewed examples for periodic SFT followed by GRPO.
SFT supplies direct supervision for the corrected SMILES-to-DSL mapping, while GRPO compares newly sampled programs on the same reaction inputs using compilation and consistency rewards.
The resulting cycle links error collection, human correction and model refinement, turning problems encountered during use into reusable training material.
This loop aims to improve robustness and generalization through accumulated corrections, with gains to be assessed in subsequent update cycles.

\section{Reaction-Grounded Constraint Compilation}
\label{sec:method}

Supervision and reward optimization left avoidable errors in fields already determined by the reaction input.
We therefore optimize the admissible generation space through \emph{reaction-grounded constraint compilation}: derive fixed facts and allowed action branches from the supplied endpoints, encode them in a per-input schema, and enforce that schema during generation.
The frozen model selects and arranges actions within this space; execution checks then test their joint consistency.
This design changes the inference interface without updating model weights.

\subsection{Compiling Endpoint Facts}

Prompt instructions alone leave known chemical facts subject to generation errors.
For a mapped reaction $x=(G_R,G_P)$, RDKit~\cite{RDKit2025} parses both endpoints, and the binding stage checks map uniqueness on each side, element correspondence, atom inventory and total charge.
The constraint constructor $S_x=\mathcal{C}(G_R,G_P)$ encodes molecular formulas, initial graph records and product memberships using JSON Schema constants and enumerated branches.
It uses the supplied reaction endpoints, without a reference DSL or a predicted edit sequence.
The same bindings provide the reference for endpoint verification after execution.

\subsection{From Program Shape to Action Restrictions}

\begin{figure*}[t]
\centering
\includegraphics[width=.98\textwidth,height=.28\textheight,keepaspectratio]{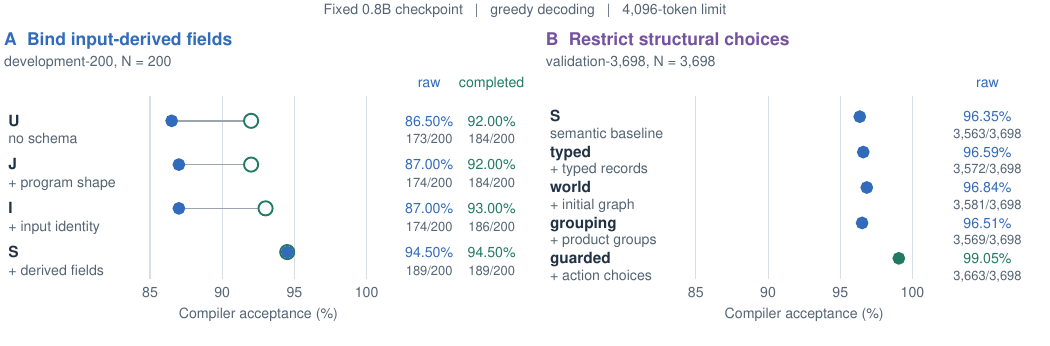}
\caption{\textbf{Compiler acceptance under each constraint family.}
(A) Binding input-derived fields on development-200: filled dots are raw generations, rings the same outputs after field completion.
(B) Cumulative structural restrictions on validation-3,698, ending in the deployed \code{guarded} configuration (green).
Both panels use one fixed 0.8B checkpoint and count failed outputs in every denominator; they use different case sets and are not one curve.}
\label{fig:constraints}
\end{figure*}

Well-formed JSON can still misstate the input, so the first constraint family separates program shape, identity binding and derived facts (\Cref{fig:constraints}A).
\textbf{U} uses the shared prompt without a schema; \textbf{J} constrains the DSL root, version, object layouts, action types and single-event structure.
\textbf{I} additionally fixes the input string and reactant names, restricts atom identifiers with their elements, and constrains atom and regroup membership choices and array lengths.
\textbf{S} (\code{semantic\_v1}) further fixes formulas, equation text and reactant/product terms.
I and S still permit repeated members, and S does not fix the complete initial bond and charge records.

Derived-field constraints still leave structural errors, motivating the cumulative second family (\Cref{fig:constraints}B).
Starting from S, \code{typed} couples bond order and type and constrains duration and identifier formats; \code{world} fixes the complete initial molecular state; \code{grouping} fixes each product's name and member list.
Finally, \code{guarded} allows break/order actions only on initial bonds, with an order change differing from the initial value, and form actions only between distinct, initially unbonded atoms.
Charge updates are limited to atoms whose endpoint charges differ, with their product charges fixed; focus targets must name known atoms, and either orientation of an undirected bond is accepted.
The model still chooses which allowed actions to emit and their order: an allowed new bond need not occur in the product graph.

\subsection{Token Masking and PICARD}

\begin{figure*}[t]
\centering
\includegraphics[width=.98\textwidth,height=.30\textheight,keepaspectratio]{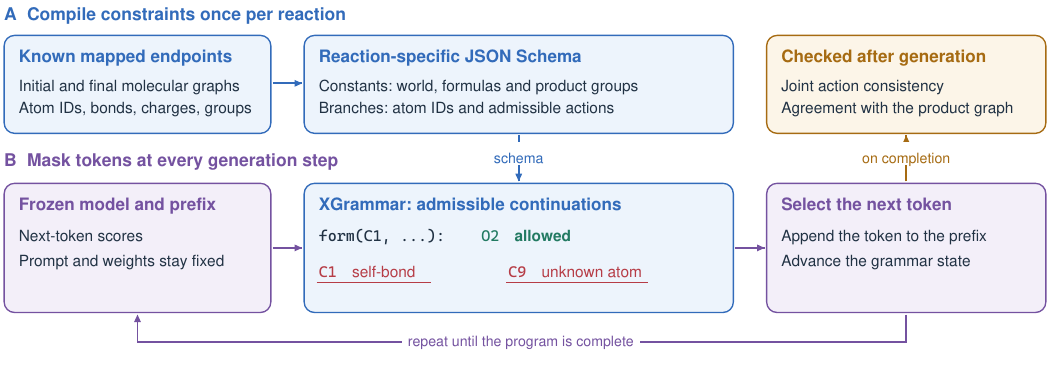}
\caption{\textbf{From reaction facts to constrained token selection.}
(A) The known endpoints define a per-reaction schema before generation.
(B) At each step, XGrammar masks incompatible continuations, the model selects a token, and the prefix and grammar state advance.
The illustrative SN2 choices reject a self-bond and an unknown atom; each identifier may span multiple tokens.
Chemical execution and endpoint agreement are checked on the completed program.}
\label{fig:decoding-mechanism}
\end{figure*}

Checking only completed outputs cannot prevent an invalid prefix from being generated.
We pass $S_x$ through the inference engine's \code{structured\_outputs} interface to XGrammar~\cite{dongXGrammarFlexibleEfficient2024}, which masks disallowed tokens before selection (\Cref{fig:decoding-mechanism}).
At prefix $h_t$, decoding uses $p_{\mathrm{con}}(v\mid h_t,x)\propto p_\theta(v\mid h_t,x)\mathbf{1}[v\in V(S_x,h_t)]$, where $V$ contains admissible next tokens.
The grammar state advances after each selected token; it does not execute a chemical graph update at each step.
The paired arms keep the prompt and model weights fixed and vary the schema, using vLLM~\cite{Kwon2023vLLM} for batch evaluation.

Incremental validity checking is established by PICARD~\cite{Scholak2021PICARD}, whose SQL parser rejects incompatible continuations using database-schema information and guards on names, aliases and scope.
Our distinction is the reaction-specific constraint construction: both chemical endpoints are known, so complete initial-state records, product groups and allowed edit branches can be compiled before decoding.
We implement this construction with XGrammar rather than a custom incremental SQL parser.
Our method builds on these established decoding principles through a chemistry-specific allocation of fixed facts, model choices and execution checks: the object of optimization is the reaction-conditioned program space.

\subsection{Completion and Execution Verification}

Post-generation field completion can recover some errors without constraining decoding, so we retain it as a separate comparison.
After checking protected input, identity, graph and event structure, it may rewrite only formulas, reactant/product terms and equation fields; it cannot change actions, atom identities, bonds or charges.
Edits are logged, endpoints are rechecked, and refusal preserves the raw candidate.
Strict post requires successful completion, compiler and endpoint checks and a normal stop; strict raw additionally requires an empty completion patch.
For an accepted S-conforming plan, the permitted derived-field patches are already empty.

Our static action restrictions still admit repeated writes, missing edits and incomplete product coverage.
They are constructed from the initial and final graphs once, rather than recomputed from a simulated chemical state after every generated action.
These are limits of the implemented schema, not a claim that JSON Schema cannot express uniqueness.
Global action consistency and agreement with the product graph remain the execution verifier's responsibility.

The compiler resolves references, checks action preconditions, valence and conservation, and returns acceptance or a typed diagnostic.
The pipeline separately verifies the executed endpoint against the supplied products before materialization and frame sampling.
Compiler acceptance in the experiments measures the compiler stage; it does not by itself certify endpoint agreement, sampled geometry or visual quality.

\subsection{Browser Realization}
\label{sec:runtime}

An accepted program still needs geometry and playback to become a usable visualization.
The public ChemQwen3D application connects mapped reactant/product SMILES to an interactive 3D scene while keeping the generated plan accessible; a stored SN2 example is also available.

The exported 0.8B student runs through ONNX Runtime and WebGPU, RDKit/WebAssembly constructs input bindings, XGrammar/WebAssembly enforces the schema, and a playback worker performs execution, endpoint and geometry checks.
This browser integration demonstrates the complete application path; the quantitative comparisons in \Cref{sec:experiments} use the batch implementation.

\section{Experiments}
\label{sec:experiments}

We ask two questions: can a 0.8B model generate dependable animation plans, and which parts of the framework are responsible.
The first is answered by a paired comparison on 19,028 reactions, the second by the two constraint families, an analysis of what exact match measures, and a breakdown of the remaining failures.
Every decoding comparison uses the same frozen \code{distilled} checkpoint with greedy decoding, temperature 0, seed 20260904, a 4,096-token output budget and thinking disabled; only the schema changes between arms.

\paragraph{Metrics and provenance.}
\textbf{Parse} counts outputs that are complete JSON documents, and \textbf{compiler acceptance} counts plans accepted by the execution checks of \Cref{sec:method}.
\textbf{Full match} compares the parsed plan with the reference as JSON objects, ignoring key order and whitespace but keeping array order; \textbf{reaction match} applies the same comparison to the \code{reaction} subtree alone; \textbf{normalized match} first applies the whitelist of order-insensitive containers defined in \Cref{sec:contract}.
\textbf{Strict raw} and \textbf{strict post} are the pipeline predicates of \Cref{sec:method}.
Parse and match counts are recomputed from the archived raw outputs; compiler verdicts are recounted from archived flags rather than re-executed; strict counts come from the pipeline reports, whose full execution logs are not part of the export, and are labeled as reported wherever they appear.
Every denominator includes failed outputs, and paired differences use exact McNemar tests on discordant cases.

\subsection{Main Result on 19,028 Paired Inputs}
\label{sec:main-result}

On the same 19,028 inputs (\Cref{tab:main}), \code{guarded} raises compiler acceptance from 18,321 (96.28\%) to 18,896 (99.31\%): 660 cases are gained and 85 lost, a net 575 cases or 3.02 percentage points ($p=4.09\times10^{-111}$).
Compiler failures fall from 707 to 132.
Every \code{semantic\_v1} output parses; seven \code{guarded} outputs do not.

Program agreement tells a more specific story.
Normalized full match rises from 16,728 (87.91\%) to 17,901 (94.08\%), yet normalized reaction match is essentially unchanged, 17,912 against 17,901, with 101 cases matching only under guarded and 112 only under semantic ($p=0.493$).
The full-match gain therefore comes from the world, not from the reaction: 1,184 semantic plans match the reaction subtree but misstate the initial world, guarded removes exactly that error while losing 11 reaction matches ($17{,}901-16{,}728=1{,}184-11$), and the model's choice of transformation shows no detectable difference between the arms.
The reported strict raw counts are 16,744 and 17,967, and on the earlier pilot-200 compiler acceptance is 189 against 195.

\begin{table}[t]
\centering
\caption{Paired comparison on 19,028 inputs with one frozen checkpoint.
Compiler verdicts are recounted from archived flags; parse and match counts are recomputed from raw outputs.
Normalized rows use whitelist C of \Cref{sec:contract}; strict raw is a pipeline-reported result.
All denominators are 19,028.}
\label{tab:main}
\small
\begin{tabular}{lrr}
\toprule
Metric & Semantic & Guarded \\
\midrule
Parse & 19,028 & 19,021 \\
Compiler acceptance & 18,321 & 18,896 \\
\quad percentage & 96.28\% & 99.31\% \\
\quad paired gain/loss & \multicolumn{2}{c}{$+660$ / $-85$} \\
Normalized full (C) & 16,728 & 17,901 \\
\quad percentage & 87.91\% & 94.08\% \\
Normalized reaction (C) & 17,912 & 17,901 \\
Normalized world (C) & 17,349 & 19,021 \\
\midrule
Original full match & 16,438 & 0 \\
Original reaction match & 17,912 & 0 \\
\midrule
Strict raw (reported) & 16,744 & 17,967 \\
\bottomrule
\end{tabular}
\end{table}

\subsection{Why the Constraints Improve Reliability}
\label{sec:mechanism}

\textbf{Binding derived fields fixes raw generation, and repair only partly substitutes for it.}
On development-200 all four arms parse 200/200, but compiler acceptance is 86.50\%, 87.00\%, 87.00\% and 94.50\% for U, J, I and S (\Cref{fig:constraints}A); program shape and identity constraints alone barely move the number, and fixing the derived fields is the step that matters, adding 15 accepted cases from I to S.
Field completion recovers part of this gap for the weaker schemas, lifting U, J and I to 92.00\%, 92.00\% and 93.00\%, while S stays at 94.50\% because there is nothing left to complete; the reported strict raw and post counts, 158/159/159/180 and 177/177/179/180, follow the same pattern.
For reference, the training report gives the deterministic endpoint-difference projector 199 compiler accepts and 198 strict endpoint passes on the same 200 inputs; it is the rule-based implementation of the same interface, and the learned arms measure how closely a fixed model approaches it.

\textbf{Restricting the admissible actions gives the largest gain.}
On validation-3,698, semantic, typed, world, grouping and guarded accept 96.35\%, 96.59\%, 96.84\%, 96.51\% and 99.05\% (\Cref{fig:constraints}B); the intermediate tiers are not monotonic, grouping losing 12 cases relative to world, and the final step of enumerating the admissible actions adds 94 accepted cases over grouping and 125 gains against 25 losses over semantic.
The gain concentrates in addition (301 to 350 of 353) and substitution (2,262 to 2,299 of 2,326), with smaller gains in condensation (780 to 790 of 793) and cycloaddition (19 to 23 of 25); a train-side check on 256 cases gives 252, 252, 252, 251 and 251 across the five tiers.
An earlier external U-to-S comparison confirms the derived-field effect at scale, raising compiler acceptance from 3,240 to 3,562 on validation-3,698 and from 13,201 to 14,188 on test-14,800; its S run and the structural-family semantic run are distinct runs of the same schema (3,562 versus 3,563 accepts).

\subsection{What Exact Match Measures}
\label{sec:contract}

Exact match on this batch depends on serialization order, and the dependence produces a striking artifact.
The guarded schema writes initial atom arrays and regroup member arrays as constants in ASCII identifier order, whereas the reference plans list atoms in map-number order and regroup members in their own order; \code{[C2,C10]} in a reference becomes \code{[C10,C2]} under the schema.
Of 42,286 reference regroup actions, 10,836 happen to be ASCII-ordered, but no case has every regroup list in that order, so no complete guarded output can equal its reference under an order-sensitive comparison, and the original full and reaction match of guarded are exactly zero (\Cref{tab:main}).
This is a mismatch between two serialization conventions, not evidence about the model; aligning the schema constants with the reference serialization would remove it.

We therefore report a sensitivity analysis under an explicit normalization whitelist: level A sorts initial atom arrays and regroup members, B also sorts initial bond arrays, and C also canonicalizes the endpoints of each undirected \code{between} bond, while actions and every other array keep their order.
For \code{semantic\_v1}, levels A, B and C give 16,447, 16,621 and 16,728 full matches against 16,438 originally; for \code{guarded}, level C gives 17,901 full and reaction matches and 19,021 world matches, and one further case would match if two \code{break} actions were swapped.
On validation-3,698 the same mechanism appears case by case: original full match falls from 3,050 for semantic to 110 for world and 20 for grouping and guarded, and reaction match from 3,344 to 30, and these are exactly the references whose atom arrays, or atom and regroup arrays, happen to be ASCII-compatible.

\subsection{Remaining Failures}
\label{sec:failures}

The 132 guarded compiler failures fall under four overlapping diagnostics: unparseable output (7), repeated regroup (70), repeated bond write (55) and repeated charge write (13); no accepted guarded output carries any of them.
All but the first are of one kind: the schema admits each action individually but says nothing about the set, so a plan can regroup an atom twice or write the same bond in both of its spellings, and only execution detects the conflict.
The semantic failures are more diverse, with 585 of 707 outside these four labels, and they include an error the compiler does not see: 1,677 semantic plans disagree with the reference initial-bond multiset and 1,260 of those are nonetheless accepted, because a plan can be internally consistent while starting from the wrong molecule.
Fixing the world at generation time is what removes this class.

The results thus divide the labor: the 0.8B model assembles the transformation, input-derived constants make the initial state right, and action restrictions remove most compiler rejections; what remains points to tracking emitted actions during decoding so that the set, not only each member, is constrained.

\section{Discussion and Limitations}
\label{sec:discussion}

\paragraph{Scope.}
The task we evaluate is single-event animation planning from known, mapped endpoints.
Compiler acceptance measures whether a plan survives the implemented execution checks and reference matching measures agreement with constructed plans; neither judges chemical mechanism, geometric realism or instructional value, which remain to be evaluated separately.

\paragraph{Model size.}
The 0.8B model runs in a browser, but we do not claim that it is cheaper or faster: the constraint studies hold model size and decoding budget fixed but do not measure latency, peak memory, throughput or energy, and the quadratic growth of the \code{guarded} enumeration makes those measurements worth taking across reaction sizes.

\paragraph{Evaluation contracts.}
The ordering analysis carries a general lesson: whenever exact match treats array order as meaningful, the constants a schema writes and the serialization a reference uses must agree, or a perfect model scores zero.
The repeated-write failures suggest the next constraint: tracking emitted action identities during decoding.

\paragraph{Data and next steps.}
Our evidence comes from one constructed corpus, one checkpoint and greedy decoding; recorded groups capture taxonomy, endpoint templates or reaction identity, but shared construction patterns can induce dependence beyond those keys, and a full replay of training would additionally require the consumed-example manifests.
Natural next steps are to vary construction coverage and model size, and to relax the input: partially mapped structures, and eventually natural-language reaction descriptions, would move the task from animating a known change toward proposing one.

\paragraph{Conclusion.}
\system turns a reaction equation into an interactive 3D animation by way of a program: a small language describes the molecular world and the change to show, a learned generator writes it, and an execution runtime checks and renders it.
A constructed corpus of 34,294 programs makes the translation learnable at 0.8B scale, and compiling the facts a reaction already fixes into per-input decoding constraints raises compiler acceptance to 99.31\% on 19,028 paired inputs without touching the model's weights.
The released browser application runs the whole pipeline on the client, from mapped SMILES to a scene in which every atom keeps its name.

{
  \small
  \bibliographystyle{ieeenat_fullname}
  \bibliography{refs}

\begin{thebibliography}{25}
\providecommand{\natexlab}[1]{#1}
\providecommand{\url}[1]{\texttt{#1}}
\expandafter\ifx\csname urlstyle\endcsname\relax
  \providecommand{\doi}[1]{doi: #1}\else
  \providecommand{\doi}{doi: \begingroup \urlstyle{rm}\Url}\fi

\bibitem[Dong et~al.(2024)Dong, Ruan, Cai, Lai, Xu, Zhao, and
  Chen]{dongXGrammarFlexibleEfficient2024}
Yixin Dong, Charlie~F. Ruan, Yaxing Cai, Ruihang Lai, Ziyi Xu, Yilong Zhao, and
  Tianqi Chen.
\newblock {{XGrammar}}: {{Flexible}} and {{Efficient Structured Generation
  Engine}} for {{Large Language Models}}, 2024.

\bibitem[Hanson(2010)]{Hanson2010Jmol}
Robert~M. Hanson.
\newblock Jmol -- a paradigm shift in crystallographic visualization.
\newblock \emph{Journal of Applied Crystallography}, 43\penalty0 (5):\penalty0
  1250--1260, 2010.

\bibitem[Hu et~al.(2022)Hu, Shen, Wallis, {Allen-Zhu}, Li, Wang, Wang, and
  Chen]{Hu2022LoRA}
Edward~J. Hu, Yelong Shen, Phillip Wallis, Zeyuan {Allen-Zhu}, Yuanzhi Li,
  Shean Wang, Lu Wang, and Weizhu Chen.
\newblock {{LoRA}}: {{Low-rank}} adaptation of large language models.
\newblock In \emph{International Conference on Learning Representations}, 2022.

\bibitem[Humphrey et~al.(1996)Humphrey, Dalke, and Schulten]{Humphrey1996VMD}
William Humphrey, Andrew Dalke, and Klaus Schulten.
\newblock {{VMD}}: {{Visual}} molecular dynamics.
\newblock \emph{Journal of Molecular Graphics}, 14\penalty0 (1):\penalty0
  33--38, 1996.

\bibitem[Jiang et~al.(2026)Jiang, Au, Xiang, and
  Chen]{jiangLaMoGenLanguageMotion2026}
Junkun Jiang, Ho~Yin Au, Jingyu Xiang, and Jie Chen.
\newblock {{LaMoGen}}: {{Language}} to {{Motion Generation Through LLM-Guided
  Symbolic Inference}}, 2026.

\bibitem[Jin et~al.(2017)Jin, Coley, Barzilay, and Jaakkola]{Jin2017WLDN}
Wengong Jin, Connor~W. Coley, Regina Barzilay, and Tommi Jaakkola.
\newblock Predicting organic reaction outcomes with {{Weisfeiler-Lehman}}
  network.
\newblock In \emph{Advances in Neural Information Processing Systems}, 2017.

\bibitem[Kozma and Russell(1997)]{Kozma1997Multimedia}
Robert~B. Kozma and Joel Russell.
\newblock Multimedia and understanding: {{Expert}} and novice responses to
  different representations of chemical phenomena.
\newblock \emph{Journal of Research in Science Teaching}, 34\penalty0
  (9):\penalty0 949--968, 1997.

\bibitem[Krenn et~al.(2020)Krenn, H{\"a}se, Nigam, Friederich, and
  {Aspuru-Guzik}]{Krenn2020SELFIES}
Mario Krenn, Florian H{\"a}se, AkshatKumar Nigam, Pascal Friederich, and Alan
  {Aspuru-Guzik}.
\newblock Self-referencing embedded strings ({{SELFIES}}): A 100\% robust
  molecular string representation.
\newblock \emph{Machine Learning: Science and Technology}, 1\penalty0
  (4):\penalty0 045024, 2020.

\bibitem[Kwon et~al.(2023)Kwon, Li, Zhuang, Sheng, Zheng, Yu, Gonzalez, Zhang,
  and Stoica]{Kwon2023vLLM}
Woosuk Kwon, Zhuohan Li, Siyuan Zhuang, Ying Sheng, Lianmin Zheng, Cody~Hao Yu,
  Joseph~E. Gonzalez, Hao Zhang, and Ion Stoica.
\newblock Efficient memory management for large language model serving with
  {{PagedAttention}}.
\newblock In \emph{{{ACM}} Symposium on Operating Systems Principles}, 2023.

\bibitem[Landrum et~al.(2025)]{RDKit2025}
Greg Landrum et~al.
\newblock {{RDKit}}: {{Open-source}} cheminformatics software, 2025.

\bibitem[Pettersen et~al.(2021)Pettersen, Goddard, Huang, Meng, Couch, Croll,
  Morris, and Ferrin]{Pettersen2021ChimeraX}
Eric~F. Pettersen, Thomas~D. Goddard, Conrad~C. Huang, Elaine~C. Meng,
  Gregory~S. Couch, Tristan~I. Croll, John~H. Morris, and Thomas~E. Ferrin.
\newblock {{UCSF ChimeraX}}: {{Structure}} visualization for researchers,
  educators, and developers.
\newblock \emph{Protein Science}, 30\penalty0 (1):\penalty0 70--82, 2021.

\bibitem[Poesia et~al.(2022)Poesia, Polozov, Le, Tiwari, Soares, Meek, and
  Gulwani]{Poesia2022Synchromesh}
Gabriel Poesia, Oleksandr Polozov, Vu Le, Ashish Tiwari, Gustavo Soares,
  Christopher Meek, and Sumit Gulwani.
\newblock Synchromesh: {{Reliable}} code generation from pre-trained language
  models.
\newblock In \emph{International Conference on Learning Representations}, 2022.

\bibitem[{Qwen Team}(2026{\natexlab{a}})]{Qwen38ModelCard}
{Qwen Team}.
\newblock Qwen3.8-{{27B}} model card.
\newblock Hugging Face model repository, 2026{\natexlab{a}}.

\bibitem[{Qwen Team}(2026{\natexlab{b}})]{qwenteamQwen3508BModelCard2026}
{Qwen Team}.
\newblock Qwen3.5-0.{{8B Model Card}}.
\newblock https://huggingface.co/Qwen/Qwen3.5-0.8B, 2026{\natexlab{b}}.

\bibitem[Rego and Koes(2015)]{Rego2015ThreeDmol}
Nicholas Rego and David Koes.
\newblock {{3Dmol}}.js: {{Molecular}} visualization with {{WebGL}}.
\newblock \emph{Bioinformatics (Oxford, England)}, 31\penalty0 (8):\penalty0
  1322--1324, 2015.

\bibitem[Scholak et~al.(2021)Scholak, Schucher, and
  Bahdanau]{Scholak2021PICARD}
Torsten Scholak, Nathan Schucher, and Dzmitry Bahdanau.
\newblock {{PICARD}}: {{Parsing}} incrementally for constrained auto-regressive
  decoding from language models.
\newblock In \emph{Conference on Empirical Methods in Natural Language
  Processing}, 2021.

\bibitem[Schwaller et~al.(2019)Schwaller, Laino, Gaudin, Bolgar, Hunter, Bekas,
  and Lee]{Schwaller2019MolecularTransformer}
Philippe Schwaller, Teodoro Laino, Theophile Gaudin, Peter Bolgar,
  Christopher~A. Hunter, Costas Bekas, and Alpha~A. Lee.
\newblock Molecular transformer: A model for uncertainty-calibrated chemical
  reaction prediction.
\newblock \emph{ACS Central Science}, 5\penalty0 (9):\penalty0 1572--1583,
  2019.

\bibitem[Schwaller et~al.(2021)Schwaller, Hoover, Reymond, Strobelt, and
  Laino]{Schwaller2021RXNMapper}
Philippe Schwaller, Benjamin Hoover, Jean-Louis Reymond, Hendrik Strobelt, and
  Teodoro Laino.
\newblock Extraction of organic chemistry grammar from unsupervised learning of
  chemical reactions.
\newblock \emph{Science Advances}, 7\penalty0 (15), 2021.

\bibitem[Sehnal et~al.(2021)Sehnal, Bittrich, Deshpande, Svobodova, Berka,
  Bazgier, Velankar, Burley, Koca, and Rose]{Sehnal2021Molstar}
David Sehnal, Sebastian Bittrich, Mandar Deshpande, Radka Svobodova, Karel
  Berka, Vaclav Bazgier, Sameer Velankar, Stephen~K. Burley, Jaroslav Koca, and
  Alexander~S. Rose.
\newblock Mol* {{Viewer}}: {{Modern}} web app for {{3D}} visualization and
  analysis of large biomolecular structures.
\newblock \emph{Nucleic Acids Research}, 49\penalty0 (W1):\penalty0 W431--W437,
  2021.

\bibitem[Shao et~al.(2024)Shao, Wang, Zhu, Xu, Song, Bi, Zhang,
  et~al.]{Shao2024DeepSeekMath}
Zhihong Shao, Peiyi Wang, Qihao Zhu, Runxin Xu, Junxiao Song, Xiao Bi, Haowei
  Zhang, et~al.
\newblock {{DeepSeekMath}}: {{Pushing}} the limits of mathematical reasoning in
  open language models.
\newblock \emph{arXiv preprint arXiv:2402.03300}, 2024.

\bibitem[Sun et~al.(2024)Sun, Liu, Gu, Lim, Bhat, Tombari, Li, Haber, and
  Wu]{sunLayoutVLMDifferentiableOptimization2024}
Fan-Yun Sun, Weiyu Liu, Siyi Gu, Dylan Lim, Goutam Bhat, Federico Tombari,
  Manling Li, Nick Haber, and Jiajun Wu.
\newblock {{LayoutVLM}}: {{Differentiable Optimization}} of {{3D Layout}} via
  {{Vision-Language Models}}, 2024.

\bibitem[Tasker and Dalton(2006)]{Tasker2006Visualisation}
Roy Tasker and Rebecca Dalton.
\newblock Research into practice: {{Visualisation}} of the molecular world
  using animations.
\newblock \emph{Chemistry Education Research and Practice}, 7\penalty0
  (2):\penalty0 141--159, 2006.

\bibitem[Tversky et~al.(2002)Tversky, Morrison, and
  Betrancourt]{Tversky2002Animation}
Barbara Tversky, Julie~Bauer Morrison, and Mireille Betrancourt.
\newblock Animation: {{Can}} it facilitate?
\newblock \emph{International Journal of Human-Computer Studies}, 57\penalty0
  (4):\penalty0 247--262, 2002.

\bibitem[Weininger(1988)]{Weininger1988SMILES}
David Weininger.
\newblock {{SMILES}}, a chemical language and information system. 1.
  {{Introduction}} to methodology and encoding rules.
\newblock \emph{Journal of Chemical Information and Computer Sciences},
  28\penalty0 (1):\penalty0 31--36, 1988.

\bibitem[Wu et~al.(2001)Wu, Krajcik, and Soloway]{Wu2001Promoting}
Hsin-Kai Wu, Joseph~S. Krajcik, and Elliot Soloway.
\newblock Promoting understanding of chemical representations: {{Students}}'
  use of a visualization tool in the classroom.
\newblock \emph{Journal of Research in Science Teaching}, 38\penalty0
  (7):\penalty0 821--842, 2001.

\end{thebibliography}
}
\end{document}